\documentclass[conference]{IEEEtran}
\IEEEoverridecommandlockouts

\usepackage{hyperref}
\usepackage{amsmath,amssymb,amsfonts}
\usepackage{algorithmic}
\usepackage{listings}
\usepackage{graphicx}
\usepackage{textcomp}
\usepackage{xcolor}
\usepackage{cuted}

\def\BibTeX{{\rm B\kern-.05em{\sc i\kern-.025em b}\kern-.08em
    T\kern-.1667em\lower.7ex\hbox{E}\kern-.125emX}}

\newcommand{\EF}{{\sf EF}}
\newcommand{\AG}{{\sf AG}}
\newcommand{\circled}[1]{\raisebox{.5pt}{\textcircled{\raisebox{-.9pt} {#1}}}}

\begin{document}

\title{An Agent-based Architecture for AI-Enhanced Automated Testing for XR Systems, a Short Paper\thanks{This work is supported by EU Horizon 2020 
research and innovation programme, 
grant 856716 project iv4XR (Intelligent Verification/Validation for Extended Reality Based Systems) and by national funds through FCT, Fundação para a Ciência e a Tecnologia, 
project UIDB/50021/2020}
}

\author{\IEEEauthorblockN{I. S. W. B. Prasetya}
\IEEEauthorblockA{Utrecht University, the Netherlands \\
s.w.b.prasetya@uu.nl \\
Orcid: 0000-0002-3421-4635}
\and
\IEEEauthorblockN{Samira Shirzadehhajimahmood}
\IEEEauthorblockA{
{Utrecht University, the Netherlands}\\
s.shirzadehhajimahmood@uu.nl \\
Orcid: 0000-0002-5148-3685}
\and
\IEEEauthorblockN{Saba Gholizadeh Ansari}
\IEEEauthorblockA{
{Utrecht University, the Netherlands}\\
s.gholizadehansari@uu.nl \\
Orcid: 0000-0002-7135-5605}
\and
\IEEEauthorblockN{Pedro Fernandes}
\IEEEauthorblockA{
{INESC-ID and Instituto Superior Técnico, Univ. de Lisboa}\\
Portugal, pedro.fernandes@gaips.inesc-id.pt \\
Orcid: 0000-0002-2840-562X}
\and
\IEEEauthorblockN{Rui Prada}
\IEEEauthorblockA{
{INESC-ID and Instituto Superior Técnico, Univ. de Lisboa}\\
Portugal, rui.prada@tecnico.ulisboa.pt \\
Orcid: 0000-0002-5370-1893}
}

\maketitle

\begin{strip} 
{\scriptsize
{\bf NOTE:} This is a preprint of an article with the same title, published the proceedings of the 1st International Workshop on Artificial Intelligence in Software Testing, which is bundled in 2021 IEEE International Conference on Software Testing, Verification and Validation Workshops (ICSTW) proceedings. The paper is published by IEEE, DOI: 10.1109/ICSTW52544.2021.00044 
}
\end{strip}

\begin{abstract}
This short paper presents an architectural overview of an agent-based framework called iv4XR for automated testing that is currently under development by an H2020 project with the same name. 
The framework's intended main use case of is testing the family of Extended Reality (XR) based systems (e.g. 3D games, VR sytems, AR systems), though the approach can indeed be adapted to target other types of interactive systems.
The framework is unique in that it is an agent-based system. Agents are inherently reactive, and therefore are arguably a natural match to deal with interactive systems. Moreover, it is also a natural vessel for mounting and combining different AI capabilities, e.g. reasoning, navigation, and learning.

\end{abstract}


\section{Introduction}

The iv4XR Framework\footnote{\url{https://github.com/iv4xr-project/aplib}} is an open source agent-based framework for automated testing of 'Extended Reality' based systems. This subfamily of interactive systems includes 3D games, 3D simulations, VR systems, and AR (Augmented Reality) systems. The domain urgently needs test automation support as manual testing is becoming very expensive and tool support is scarce (even record and replay is often not available). 
An {\em agent} is essentially a program that interacts with an environment by repetitively performing actions, either on its own initiative or as reaction to events generated by the environment. 
An agent is thus inherently a {\em reactive program}, and in this sense it is fundamentally different than e.g. a procedure or a service.  Arguably, this makes agents a more natural framework for testing interactive systems.

The iv4XR Framework is currently under development by an H2020 project with the same name. It has reached a working prototype level, and is undergoing internal piloting. To deal with the huge and fine grained interaction space of XR systems, iv4XR necessarily relies on AI. It has however its own, rather unique, perspective of how AI is to be deployed to aid software testing.

While the current interest in AI mainly focuses in  machine learning, iv4XR's main AI is agent-based AI. The agent community has long insisted that intelligent agent is AI, a position that is also supported by the AI community  \cite{wooldridge1995intelligent,MeyerAgentTech2008,russellAI}. 
Iv4XR is inspired by a popular concept of intelligent agents called {\em Belief-Desire-Intent} (BDI) \cite{herzig2017bdi,bordini2007programming,dastani20082apl}. An agent is thought to have a mental state, which includes its `beliefs' and `desires'. `Desires' are formulated as goals that the agent seeks to achieve. `Intent' represents the agent's plan towards achieving a goal; this corresponds to the concept of function or method in traditional programming. The intelligent part comes from the agent's ability to reason, e.g. through reasoning rules, about its believes and goals, to decide which goal to pursue, and which plan to use. This approach combines a {\em proactive dimension}, that together with the {\em reactive nature} of agents
provides more versatility and strength for automated testing than just invoking e.g. a random tester or a genetic algorithm. 

`Belief' is also different than the traditional concept of `state'. A BDI system acknowledges that belief is not necessarily the same as the reality. For example, suppose a test agent wants to test if a certain button in the UI of some system under test (SUT) behaves correctly. To find the button, it might {\em believe} that the button is located in e.g. $panel_x$. By acknowledging that this is only a belief, and not necessarily a fact, we would be more compelled to program what the agent should do if the belief turns out not to hold, e.g. if the developer has moved the button to $panel_y$. This mindset encourages the development of robust test strategies.

Agent-based AI and machine learning are not mutually exclusive. For example, reinforcement learning can be seen as an approach for finding a policy for an agent towards solving a goal. This can be leveraged by combining it with BDI agent's ability to reason. Reasoning rules improve the agent's ability to discern which  actions in a given state are much more likely, or else unlikely, to lead to worthy rewards towards reaching the goal, hence pruning the space that the agent has to explore to learn. 
Conversely, when not all reasoning rules are known, techniques such as rules learning can be used to learn them. 

Iv4XR has more features than just BDI, e.g. goal-based and tactical programming \cite{prasetya2020aplib}, automated navigation and exploration, and integration with other testing tools such as TESTAR \cite{vos2015testar}. A demonstration is available, where iv4XR is used to automate the testing of a 3D game called Lab Recruits\footnote{\url{https://github.com/iv4xr-project/iv4xrDemo}}. Larger pilots are work in progress.

This short paper presents an architectural overview of the iv4XR Framework.
Section \ref{sec.iv4XR} introduces the key concepts, the architecture of a test agent, and explains how it works.
Section \ref{sec.wom} explains iv4XR's design pattern to make it extensible for targeting an arbitrary SUT. 
Section \ref{sec.relatedwork} discusses some related work. Section \ref{sec.conl} concludes and mentions future work.

\section{Testing with iv4XR}\label{sec.iv4XR}

Iv4XR is implemented as a Java library. This also means that Java is the language to use when formulating testing tasks 
---the advantage is that Java is a popular language, well supported with  development tools. 
Iv4XR also comes with a set of APIs that mimic a Domain Specific Language (DSL), allowing testers to formulate tasks more fluently.
Since it is not possible to have one automated testing framework to cater all sorts of technologies and ontologies of the target systems, iv4XR is designed from the outset to be extensible, 
relying on two Java features to do this: {\em inheritance} to allow testers to customize standard behavior, and {\em $\lambda$-expression} to allow new behavior to be fluently passed as parameters.

\begin{figure}
    \begin{center}
    \includegraphics[scale=0.35]{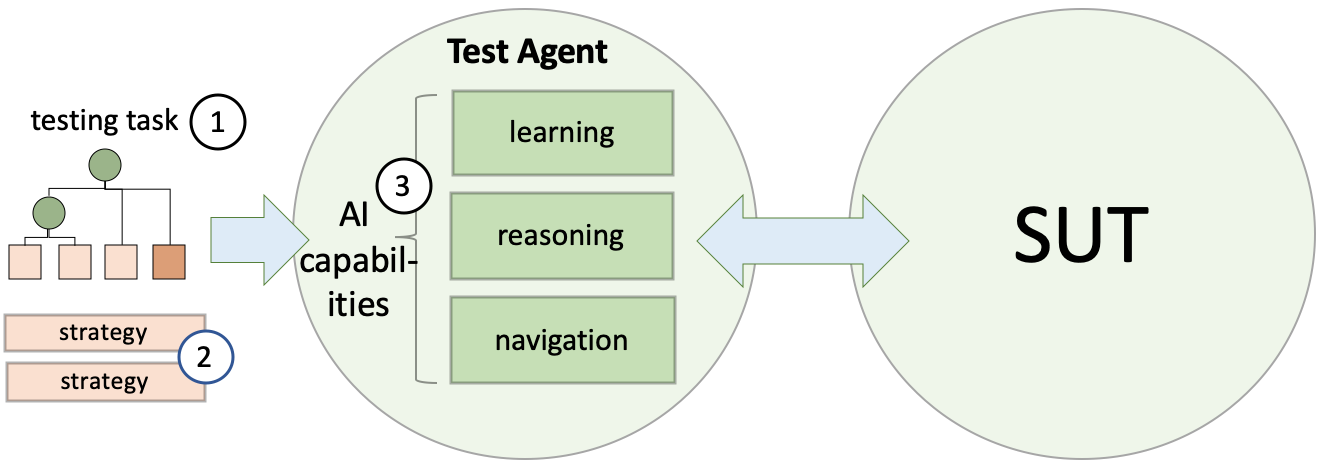}
    \end{center}
    \vspace{-4mm}
    \caption{\em An iv4XR test agent accepts testing tasks. Strategies are used to solve them, which in turn make use of general AI capabilities such as reasoning and navigation.}
    \label{fig.testagent}
\end{figure}

Figure \ref{fig.testagent} shows a top level architecture of test agents. But 
let us first explain iv4XR's concept of `automated testing'. 

\subsection{What should we automate?} \label{sec.whatshouldwe.automate}

Given a System under Test (SUT), iv4XR facilitates the development of strategies which can be used to automatically `solve' testing tasks. Strategies might be SUT-specific, but once developed they can be reused to automate all sorts of testing tasks for the SUT (or its family). While testing tasks can be automatically solved, iv4XR does expect the tester to specify what the tasks are. This is {\em different} than automation provided by tools like QuickCheck, T3, or Evosuite \cite{claessen2011quickcheck,prasetya2016budget,fraser2011evosuite} which only need the SUT to be given, after which they can generate tests. Such an approach works when testing units (e.g. a method or a class).  An interactive system is however more complicated. The state space is much larger, and has a complicated structure. Without being more specific in {\em what} we want to test, just wandering around trying different things is unlikely to be effective, which is why iv4XR is aimed towards task-level automation.

The simplest form of `testing tasks' is the task to verify an `assertion'. Two common types of assertion-like properties are these (borrowing CTL notation \cite{baier2008principles}):

\begin{itemize}
    \item {\em Existential} $\EF\phi$, asserting the existence of an execution that leads to a state satisfying $\phi$. 
    As an example: in a web application a page named $Purchase$ should be reachable.
    
    \item {\em Universal} $\AG(\phi \rightarrow \psi)$, asserting that all states satisfying $\phi$ should also satisfy $\psi$.
    For example, we might require that the aforementioned page $Purchase$ (the $\phi$ part) contains a button named $cancel$, and clicking on it should empty the user's purchasing basket (the $\psi$ part). 
    
\end{itemize}

In both cases, a testing algorithm will have to find at least one right sequence of interactions that would move the SUT to $\phi$. This part is often very challenging, whereas checking the $\psi$ part, in the case of a universal assertion, is usually straight forward. For example, if the SUT is a computer game, and $\phi$ is a key room in the game, verifying $\EF\;\phi$ would effectively require an algorithm that knows how to play the game, at least as far as getting itself to reaching the room. Obviously, this is not an easy feat.
From this perspective, `strategies' pointed in Fig. \ref{fig.testagent} are, in the heart, typically aimed to guide the agent towards different kinds of $\phi$.

\subsection{Test agent}

Figure \ref{fig.testagent} shows the architecture of a test agent. The agent controls the SUT by executing a series of {\em actions}. Actions can be expected to be very {\em primitive}. E.g. if the SUT is a computer game, an `action' could be to move an in-game character in a given direction for few frame updates. Modern games can run at the rate of 100 frame updates/second, so such an action will only move the character for a very small distance. Note that having primitive actions is a good thing, as this allows the agent to have  refined control on the SUT, though the trade off is that it needs to put more effort in planning.

To do anything, the test agent must be given a {\em testing task}, see \circled{1} in Fig. \ref{fig.testagent}, e.g. to verify an assertion as discussed above. In terms of BDI agency, a task is a {\em goal}. When the agent manages to complete the goal, we say that it is `solved'. To solve a goal, a `tactic' is needed. A tactic can be just a bunch of actions $\{ \alpha_i \}$, each can be guarded with some reasoning logic $g_i$ determining when it makes sense to execute the action. 
E.g. $\alpha_0$ could be to interact with an in-game button, and its $g_0$ could require the agent to be physically close enough to the button (else the game might not allow the interaction); $\alpha_1$ could be moving in the direction of the button, and its $g_1$ could require that the agent has a clear line of sight to the button.

Being a reactive program, the execution model of an agent is  {\em very different} than e.g. a procedure or a service.
An agent runs in (rapid) cycles. A cycle is either triggered by a clock tick or by an incoming event. At each cycle, the agent checks which actions in its current tactic are `enabled'; that is, having their logic-guard evaluates to true. One is then chosen, randomly, or according to some {\em policy}. The cycles are repeated until the goal is solved (or the agent runs out of budget). Note that the actions' guards essentially form the reasoning part for solving the goal. Iv4XR additionally provides {\em combinators} similar to the idea of tactic combinators in theorem proving \cite{delahaye2000tactic,gordon1993introduction} to structurally combine simpler tactics to construct more complicated ones. Indeed, programming a tactic might be non-trivial for testers, so this part should be made transparent, or at least mitigated. We will return to this later.

Some testing tasks can be expected to be non-trivial. To make them feasible for the test agent to solve, we can introduce subgoals. Goal-combinators can be used to specify how to combine them. The idea is similar to Behavior Tree from AI \cite{millington2009artificial}; an example is visualized below:

\begin{center}
    \includegraphics[scale=0.4]{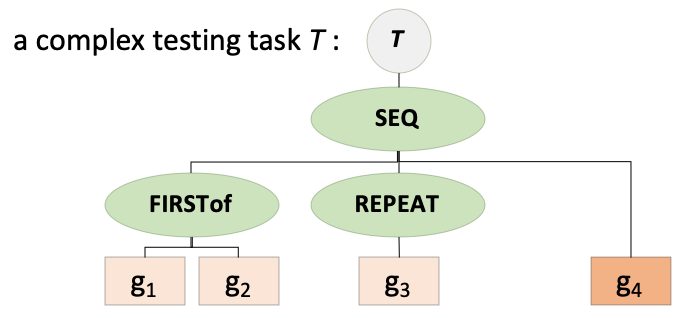}
\end{center}

E.g. the $\sf SEQ$ combinator formulates a task where all its subgoals must be solved in the specified sequence. Above, the actual testing task is $g_4$, e.g. as in a 
previous example, to verify that a key room in a game is reachable. The subgoals $g_1..g_3$ can be thought as `lemmas' to help in solving $g_4$.

Figure \ref{fig.snippet} shows an example, at the code-level, of how a testing task is specified, assigned to an agent, and how to inspect the result of the task.

Many tasks are actually just variations of each other. E.g. a testing task to check the state of some in-game button $b_0$ is just a variation of a similar task for checking another button $b_2$. We can therefore provide a {\em parameterized goal} to do this task, and instantiate it on demand whenever we need to verify some concrete button.

In Section \ref{sec.whatshouldwe.automate} we mentioned `strategies', and also as \circled{2} in Fig. \ref{fig.testagent}.
A strategy is a parameterized goal capturing a common subtask. When formulating testing tasks, we can imagine that testers have access to a library of these strategies; all they need to do is to instantiate them, and to arrange them towards solving the testing task they have in mind.

Of course, someone needs to provide the strategies in the first place. We expect them to be quite SUT specific. E.g. strategies for a shooter game cannot be expected to be reusable for a train simulator. However, constructing them is a one-off investment, after which testers can keep reusing them to write automated testing tasks.

Since a strategy is essentially a goal, we  need a tactic to solve it. So, the one-off investment also involves the development of common tactics. The basic building blocks for tactics are the primitive actions as provided by the SUT itself. To combine them, we indeed have the aforementioned tactic combinators, 
but additionally 
we also have access to a number of AI capabilities, indicated in \circled{3} in Fig. \ref{fig.testagent}:

\begin{itemize}
    \item {\em Reasoning}: through actions' guards as discussed before. 
    
    \item {\em Navigation}: steering a virtual character to navigate through a 3D virtual world is not trivial since it typically also has to respect some physical laws. E.g. the virtual character would not be able to see nor walk through a solid obstacle; this complicates navigation a lot.
    Iv4XR implements the path finding algoritm A* so that a tactic can automatically steer a test agent to a given target location. When the agent want to inspects some virtual entity, but its location it not known (the tester may deliberately abstract away the location, to make the test more robust), the agent will then have to search the entity first. Iv4XR also implements an algorithm inspired by robot exploration to do this; for more on this see \cite{prasetya2020navigation}.
    
    \item {\em Learning}: Recall that an agent uses a `policy' when determining which action among the set of enabled actions is to choose for execution. The default policy is just random, but a custom policy can be set, which in turn can be obtained through learning. 
    
    Technically, a goal is a predicate $\phi$ evaluated over a candidate $C$ proposed by the tactic $T_\phi$ associated to $\phi$. The goal is solved when $T_\phi$ finds a $C$ such that $\phi(C)$ is true. To support unsupervised learning, $\phi$ can also be formulated as a cost function: when a tactic proposes a wrong $C$, $\phi(C)$ expresses an estimation of the remaining effort, starting from what we know about $C$, to find a solution. This would allow e.g. Reinforcement Learning to be deployed to learn a policy.
    
    Note that a policy is used on {\em enabled} actions, which means it is used {\em in combination} with the agent's own reasoning (through action guards) as the latter determines which actions are enabled.
    

\end{itemize}

\begin{figure}
\begin{lstlisting}[
  mathescape=true,
  basicstyle=\scriptsize,
  numbers=left, 
  numberstyle=\tiny,
  frame=leftline,
  xleftmargin=5mm,
  keywordstyle=\color{blue},
  keywords={var,SEQ,REPEAT,FIRSTof,while,assertTrue,update,null}
  ]
var ttask = SEQ(
    FIRSTof($g_1$,$g_2$), 
    REPEAT($g_3$), 
    // $g_4$:
    assert_(agent, $\beta\rightarrow$ $\beta$.wom.getElement($key$) $\not=$ null))

agent.attachState      (new W3DAgentState.java())
     .attachEnvironment(new EnvSUT())
     .setDataCollector (new TestDataCollector())
     .setGoal          (ttask)
     .budget           (1000)
     .useDeliberation  (policy)

while(ttask.getStatus()==INPROGRESS) agent.update() ;

var verdicts = agent. getTestDataCollector() ;
assertTrue(verdicts.getNumberOfFailVerdictsSeen() == 0)
\end{lstlisting}
\caption{An sample iv4XR code. Lines 1..5 specify a testing task. The last goal, $g_4$, is a task to verify if an in-game entity $key$ is present in the game; $\beta$ represents the agent's belief.
Other goals are intermediate goals intended to guide the agent in solving $g_4$.
Line 11 assigns the aforementioned task to a test agent. Line 12 sets a computation budget for the agent, and line 13 sets an action-selection 'policy'. Line 15 runs the agent in a simple loop. Finally, line 18 checks if the agent found no violation.} \label{fig.snippet}
\end{figure}

\section{Interface Pattern and World Model} \label{sec.wom}

\begin{figure}
    \begin{center}
    \includegraphics[scale=0.32]{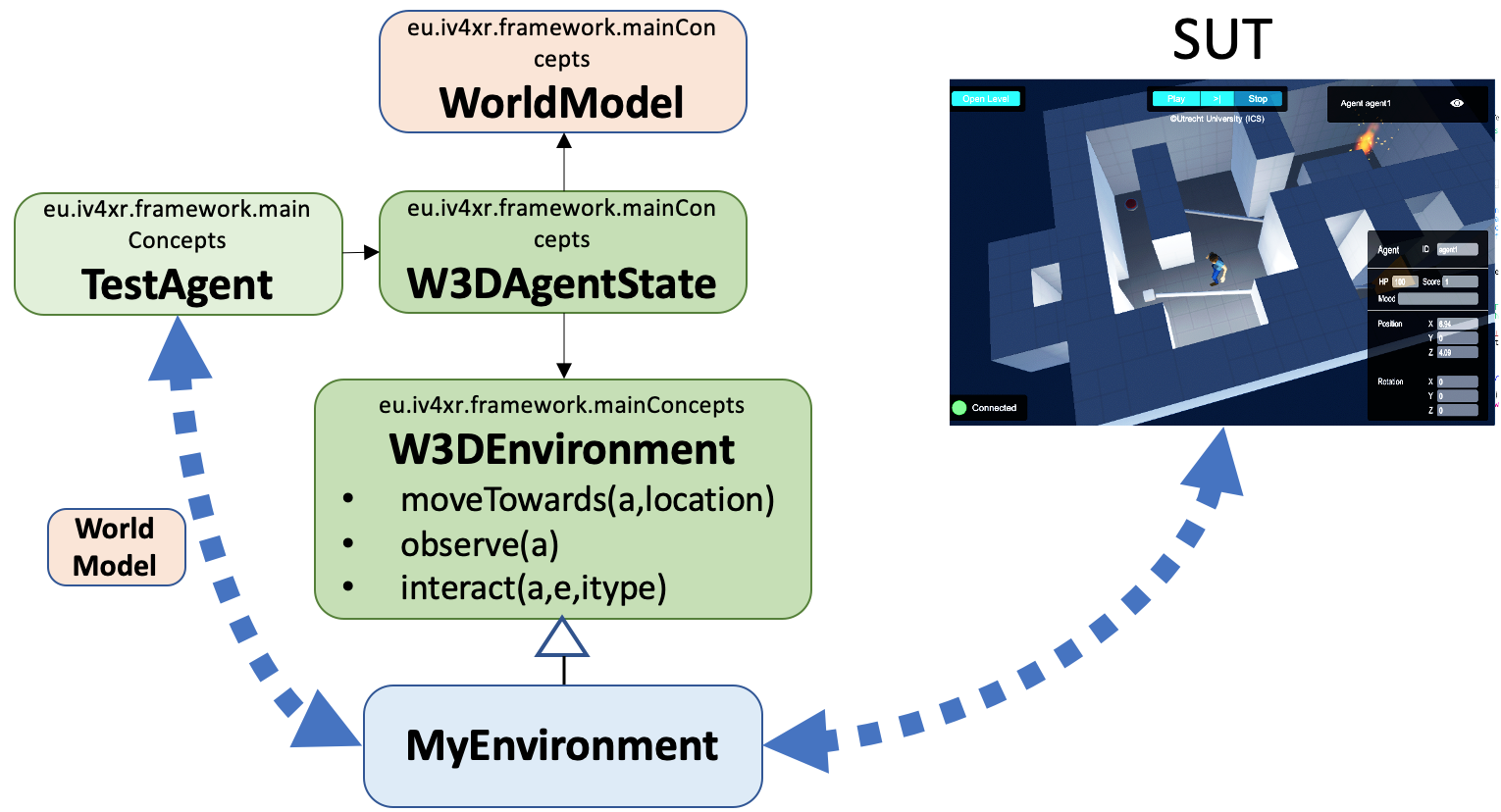}
    \end{center}
    \vspace{-4mm}
    \caption{\em The structure of the interface between iv4XR and the SUT. It is a variation of the Proxy Design Pattern \cite{gamma1995elements} with $\sf W3DEnvironment$ and $\sf MyEnvironment$ taking the role of the abstract and concrete proxies.}
    \label{fig.interface}
\end{figure}

To do its work the test agent will need an interface that lets it control the SUT and inspect its state. For XR systems, this is more challenging than e.g. browser-based applications. There is a large variation in the used UI technologies and there is no clear winner. Hence there is no common interfacing technology either. To make iv4XR usable for as much users as possible, the framework is provided open source and designed from the outset with extensibility in mind. The trade off is that a company using iv4XR first has to put some effort to instantiate its interface scheme, depicted in Figure \ref{fig.interface}.

Every test agent maintains a state ($\sf W3DAgentState$), which includes what it believes to be the current SUT state. This is represented by a so-called World Object Model (WOM), or simply World Model. The agent's state also contains a pointer to an `Environment' ($\sf W3DEnvironment$), which would provide a set of primitive actions used by the test agent to control or observe the SUT, such as to move some small distance, or to interact with an entity. The implementation of these actions is, however, SUT dependent, hence cannot be provided by $\sf W3DEnvironment$ itself. Developers therefore need to provide the implementation in the form of the class $\sf MyEnvironment$ that extends $\sf W3DEnvironment$.

Each action from the Environment will also return a WOM, representing what the agent observes at the end of the action. Being an XR system, the SUT is assumed to represent a 3D world. A WOM represents a fragment of this world. E.g. it might list relevant entities in the world, along with their key properties. While the WOM returned by an action represents what the agent {\em currently} sees, the WOM maintained in the agent state {\em aggregates} all the received WOMs. While this maximizes the information the agent memorizes, some of the information in its WOM might not reflect the actual SUT state. The information is timestamped, but it is up to the agent to decide what to do with it.

\begin{figure}
    \begin{center}
    \includegraphics[scale=0.35]{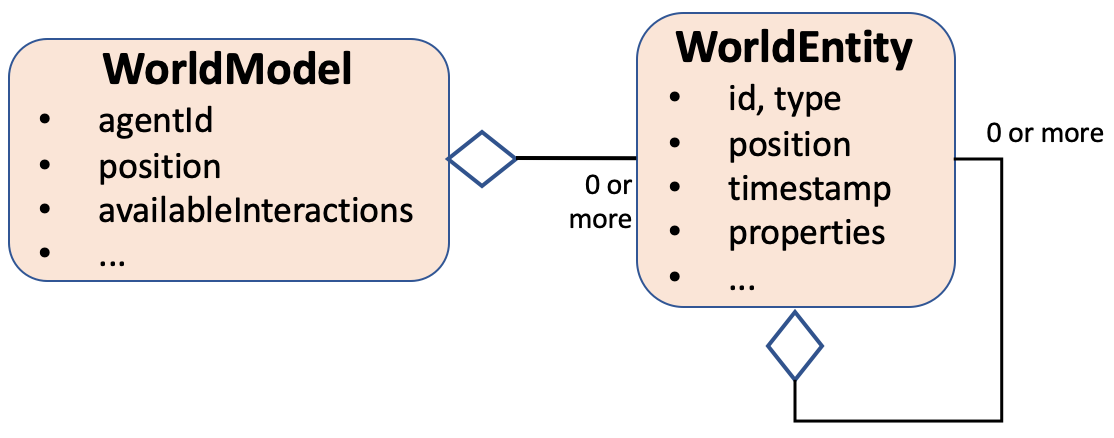}
    \end{center}
    \vspace{-4mm}
    \caption{\em The tree structure of the World Object Model (WOM). A world has 0 or more entities; each in turn may consist of multiple sub-entities. Since a WOM represents a 3D world, it keeps track e.g. the agent's last known position in the world, as well as that of the entities.}
    \label{fig.wom}
\end{figure}

Another important aspect of the WOM is that it represents a virtual world {\em structurally}, rather than visually. This allows agents to reason about the world much more accurately, which in turn also makes testing robust against all sorts of visual changes during the development; something which game designers can be expected to do quite often. Figure \ref{fig.wom} shows the structure of a WOM. It is actually inspired by Domain Object Model (DOM) that is used by browers to provide a common and structural interface to programmatically access and manipulate web pages. Importantly, a DOM is agnostic towards the actual ontology of the web page it represents. Similarly, a WOM is agnostic towards the game's specific ontology, hence allowing test strategies to be crafted more generally.


\section{Related Work} \label{sec.relatedwork}

The most studied type of AI to aid automated testing is probably Reinforcement Learning (RL). One of the early works is that of Mariani et al. \cite{mariani2014automatic} where RL is used to help learning the behavioral model of the SUT, from which test cases are then generated. RL is mainly used to train a policy that optimize the discovery of new states (in other words, to optimize the test coverage). Later works, e.g. as in the use of RL in the GUI testing tool TESTAR \cite{bauersfeld2014user}, and similarly in approaches for mobile app testing \cite{adamo2018reinforcement,vuong2018reinforcement} still follow the same idea of how RL is exploited (so, to optimize coverage). In contrast, iv4XR sees AI mainly as an instrument for solving testing tasks. 

Neural Networks (NN) have been proposed to be used to as artificial specifications \cite{vanmali2002using,aggarwal2004neural,prasetya2019neural}. The idea is to train an NN to approximate the behavior of a program $P(x)$, after which we can then use the NN as an oracle when testing $P$, in particular when the used testing algorithm generates a large amount of test cases (e.g. as in random testing). The approach suffers from false positives; even if there are only e.g. 5\%, each one will have to be manually investigated. The issue is probably not to be blamed to the NN; it is a common issue in specification learning.   

Iv4XR is not the first attempt to use agents to aid testing. Earlier works we can mention are \cite{qi2005agent,paydar2011agent,bai2011design}. These work went as far as using agents, and even BDI, but did not explore how they can actually exploit agent-based AI. 

\section{Conclusion, Current Status, and Future Work} \label{sec.conl}

We have given an overview of the iv4XR Framework. The Framework provides an agent-based approach to automatically solve testing tasks on interactive systems. It relies on agent-based AI, with possibilities to be combined with learning-based AI. The iv4XR Framework is under development and has currently reached a prototype level. Currently it is being piloted for testing 3D games, though as future work we would like to cover VR systems as well. Actual integration of learning AI is also future work, as well as the evaluation of iv4XR on more pilots.

\bibliographystyle{./bibliography/IEEEtran}
\bibliography{./bibliography/IEEEabrv,./aplibBib}

\end{document}